# CONSTRUCTION OF OBJECT BOUNDARIES FOR THE AUTOPILOT OF A SURFACE ROBOT FROM SATELLITE IMAGES USING COMPUTER VISION METHODS

**A.N. Grekov[1,2], Y.E. Shishkin[1], S.S. Peliushenko[1], A.S. Mavrin[1,2]**

[1]Institute of Natural and Technical Systems, RF, Sevastopol, Lenin St., 28

[2]Sevastopol State University, RF, Sevastopol, Universitetskaya St., 33

*E-mail: iurii.e.shishkin@gmail.com*

An algorithm and a program for detecting the boundaries of water bodies for the autopilot module of a surface robot are proposed. A method for detecting water objects on satellite maps by the method of finding a color in the HSV color space, using erosion, dilation - methods of digital image filtering is applied. The following operators for constructing contours on the image are investigated: the operators of Sobel, Roberts, Prewitt, and from them the one that detects the boundary more accurately is selected for this module. An algorithm for calculating the GPS coordinates of the contours is created. The proposed algorithm allows saving the result in a format suitable for the surface robot autopilot module.

**Introduction**. In recent years, technological development and the accelerated rate of population growth have led to an increase in anthropogenic pressure and an increase in the level of human use of natural resources, which has caused climate change. This leads to significant negative trends in changes in the state of the natural environment, hydrosphere and atmosphere. One of the key backbone elements of the ecosystem are surface water bodies: rivers, lakes, seas and oceans. Water is an important element for the existence of various ecosystems, including human existence. As a result, automated monitoring of water bodies is an important process in scientific and practical research [1].

In fact, there is currently no permanent monitoring of water bodies that would record changes in their physicochemical parameters and areas. To automate the processes of such monitoring, an effective and promising direction is the use of an unmanned surface robot [2–4]. At the same time, the task of automating the process of constructing maps for the autopilot of such a robot is not exhaustively and fully resolved, there are known attempts to solve it based on ensemble, probabilistic and convolutional algorithms [5, 6]. In our work, we solve the problem of constructing maps for the autopilot of a surface robot based on satellite images of the area using computer vision methods.

To implement the method of constructing boundaries for the autopilot, there are many different algorithms with different efficiency for unique target problems [1, 7, 8]. In this paper, we study the effectiveness of methods for constructing the contours of water bodies on satellite images using a number of operators [9]:
– is the Sobel operator,
– is the Roberts operator,
– is the Prewitt operator.

The following approaches are proposed in the literature for the detection of water bodies:
– Methods of multispectral indices – NDWI (normalized difference water index, normal different water index), MNDWI (modified normalized difference water index, modified normal different water index), AWEIsh/nsh (automated water detection index with/without shade, automated water extraction index shadow/not shadow variants), NSVDI (normalized saturation-value difference index) [1, 10].
– Morphological detection methods [11, 12].
– ML classifier detection methods [11].
– A special detection algorithm using index methods and morphological methods [12].
– SBDS method (single-band density slicing) [13].

The solution to this problem has been considered by many authors. In particular, in [11], the morphological detection method (MM) and ML classifier detection methods are studied, and the result of the analysis is compared with the results of methods based on water indices (NDWI, MNDWI, AWEIsh/nsh). As a result, MM is recognized as the best in most cases.

The paper [12] proposes an algorithm for detecting water objects in an image using water index methods (NDWI, MNDWI, and NSVDI) and morphological detection methods.

In [13], methods for detecting water bodies based on NDWI, MNDWI, AWEIsh/nsh methods of multispectral indices with the addition of methods for determining thresholds and clustering are studied to increase the accuracy of detection. The improved accuracy of the method for detecting objects with increased brightness is recognized.

In [10], an automated method for extracting rivers and lakes by combining water indices with digital image processing methods is proposed, and the higher efficiency of this solution for the problem of detecting water objects is noted compared to methods based on indices (NDWI, MNDWI, AWEIsh / nsh) due to better detection of mixed pixels and noise.

In [13], the features of the SBDS method for the Landsat TM (Thematic Mapper) image are studied. It is recognized that Landsat maps are superior to satellite imagery for feature detection, using MatLab and ENVI tools on a Landsat image and proper band selection can achieve excellent results in mapping a water body and separating it from surrounding land.

In our work, for image processing, we used the method of detecting certain colors in an HSV image with a predetermined color palette of water. Next, on the processed image, we used the methods of removing noise, erosion, dilation, and contour detection using the Sobel operator. After that, to detect the GPS coordinates of the contour, they used a specially developed algorithm described in this article.

The following software was used to develop the algorithm: SASPlanet [14] and MATLAB Rb2017 [15]. To obtain an image of a map fragment, the SASPlanet program is used, since it is freely distributed. The MATLAB program was chosen because it already implemented the functions for constructing boundaries using the Sobel operator.

**Algorithm for detecting the boundaries of water bodies.** It is a sequence of 9 stages that implement a full range of measures aimed at solving the target problem from loading satellite images and highlighting the boundaries of water bodies to saving the results in a form suitable for use in the autopilot module.

**Stage 1. Image acquisition.** The input image is a map fragment. The developed program uses a fragment of the image received by the operator from the SASPlanet program, the Yandexmap satellite section. For the algorithm to function, it is necessary to specify the target area of research. An example is shown in fig. one.

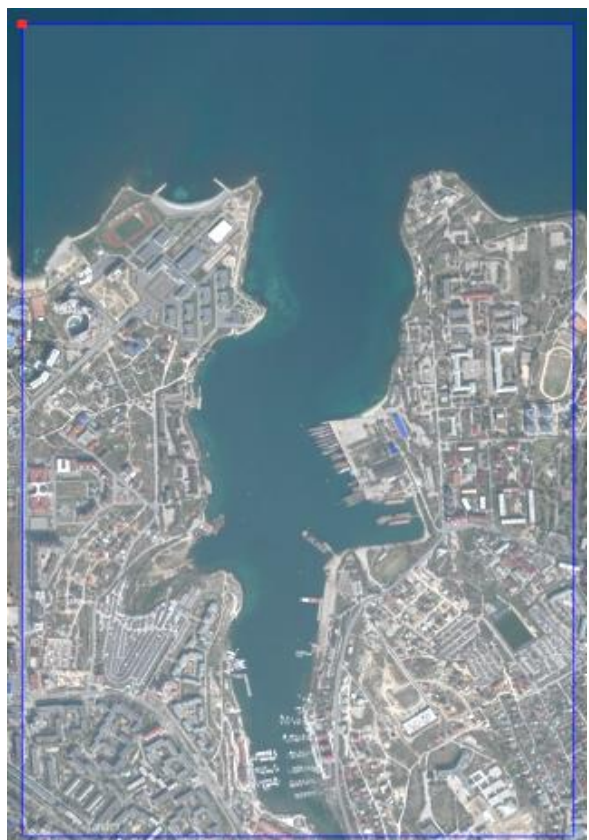

Fig. 1. Map fragment. The red point is the starting point of the selection. The highlighted area is the initial example of the execution of the algorithm under study

**Stage 2.** Implementation of image binarization. The original image is converted into a binary image, the elements of which can take only two values - binarization.

A distinctive feature of this algorithm is the conversion of the image from the RGB color model to the HSV color model before binarization.

To convert RGB to HSV, relations (1)–(3) are used:

$$H = \begin{cases} 60 \times \frac{G-B}{MAX-MIN} + 0, \text{если } MAX = R \text{ и } G \geq B \\ 60 \times \frac{G-B}{MAX-MIN} + 360, \text{если } MAX = R \text{ и } G \leq B \\ 60 \times \frac{B-R}{MAX-MIN} + 120, \text{если } MAX = G \\ 60 \times \frac{R-G}{MAX-MIN} + 240, \text{если } MAX = B \end{cases} \quad (1)$$

$$S = \begin{cases} 0 \text{ если } MAX = 0; \\ 1 - \frac{MIN}{MAX}, \text{иначе} \end{cases} \quad (2)$$

$$V = MAX \quad (3)$$

where R, G, B - red, green, blue - parameters of the RGB model (a color model that describes the method of color coding for color reproduction using three colors, which are commonly called primary: red, green, blue - red, green, blue); HSV - hue, saturation, value - parameters of the HSV model (color model, in which the color coordinates are: hue - hue, saturation - saturation (the larger this parameter, the more saturated the color, and the closer this parameter to zero, the closer the color to neutral gray), value – color value).

$H \in [0,360], S, V, R, G, B, \in [0,1]$, Max is the maximum value from R, G, and B, MIN is the minimum value, which is undefined if MAX=MIN.

In the MatLab environment, this translation was made by the rgb2hsv function.

After receiving an image in the HSV model, its mask is built - an image, each pixel of which can have only one of two values: 0 (logical false) and 1 (logical true); in the image, 0 is plotted as black and 1 as white. It is worth noting that the original array describing the image consists of 3 columns - describing 3 coordinates of the HSV color model of each pixel.

The array describing the mask consists of 1 column - describing its true/false. To build a mask, you need a condition for which each pixel of the source image will be checked, and if the pixel meets the condition in the mask, it will be equal to 1, and if it does not, it will be equal to 0.

Since the task is to find water objects, the condition is 3 ranges of values for each of the coordinates of the color model, which together describe the color range of water (these ranges were selected using the MatLab subroutine - Color Threshholder):

– Hue value range [0.399:0.78] (133° – 260°);

– Saturation value range [0.32:1] (32% – 100%);

– value range [0.2:1] (20% – 100%).

Therefore, to build an image mask, it is necessary to check the coordinates of each pixel for belonging to these ranges (i.e., for each pixel, a check is made: if all three pixel coordinates belong to the corresponding ranges, then the element corresponding to the pixel in the array describing the image mask is assigned 1, otherwise 0). The result is a binary image, where water objects are highlighted in white, and everything else is in black.

On fig. 2 you can see the result of the method - applying a mask to the original image. Using the figure, you can check the correctness of the detection of water objects by choosing colors.

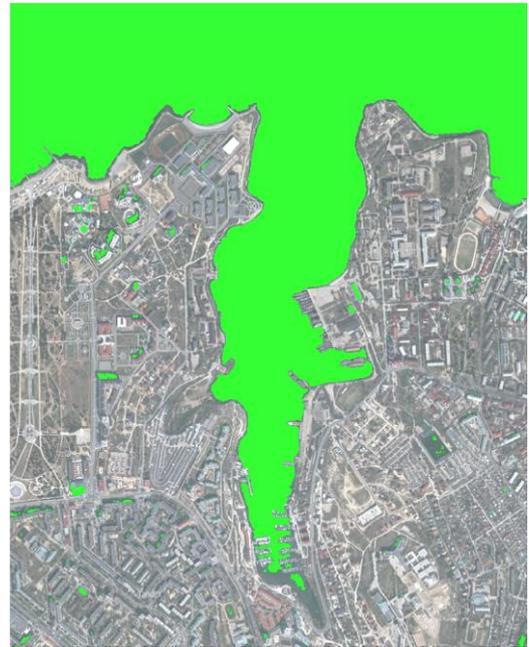

**Fig. 2.** The result of the work of the color selection method in the HSV color space

**Stage 3.** Removal of small areas. The dilation operation is aimed at increasing the adequacy of the selection of an image area by expanding its pixels and thereby facilitating the unification of image areas that have been separated as a result of noise, etc.

The image after filtering becomes lighter and visually more blurred. That is, dark details are weakened or disappear altogether, which depends on the ratio of their size and brightness with the given filter parameters [16].

This is the operation of determining a local maximum in a certain neighborhood, which is specified by a structure-forming element. The action of the dilation operation in relation to the example under study can be seen in Fig. 3.

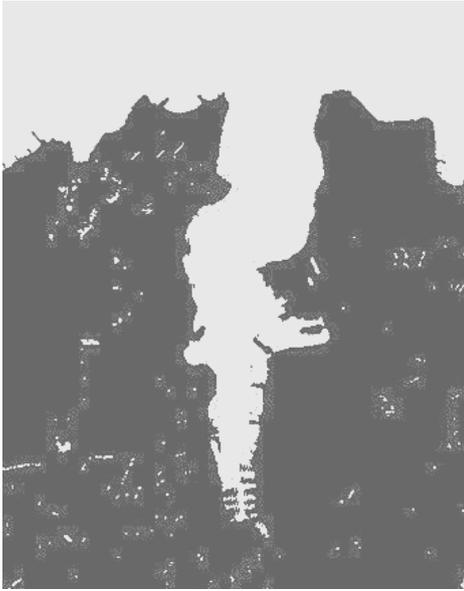

**Fig. 3.** Application of the dilatation operation

The erosion operation is one of the main operations of mathematical morphology along with dilation transformations. This is the reverse operation of dilatation. Erosion is used to reduce an area of an image, resulting in thinning of pixels, widening and enhancing highlights in an image.

The essence of this transformation is that unwanted inclusions and noise are blurred, and large and, accordingly, significant areas of the image are not subject to changes.
Unlike dilation, this operation defines a local minimum in some neighborhood, which is given by a structure-forming element.

Let a structural element B be given. The erosion of the set X is the set Y, which consists of those elements of the original set X for which the condition

$$Y = X \ominus B = \{x: B_x \leq X\} \quad (4)$$

In other words, if x ∈ X, b ∈ B, then the set includes such elements x for which, for all b, the condition

$$x + x \in b \quad (5)$$

Erosion can be interpreted as follows: we place the center of the structural element at the point x ∈ X, if the element completely belongs to X, then the point x ∈ Y, where is the set of similar points. Obviously, after this operation, the set Y is less than X. The effect of the operation is illustrated in fig. 4 [16].

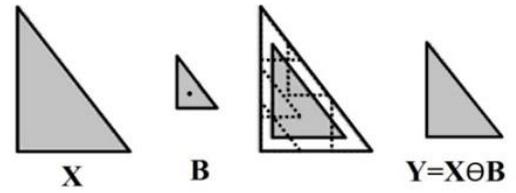

**Fig. 4.** An example of an erosion operation

The action of the erosion operation in relation to the example under study can be seen in Fig. 5.

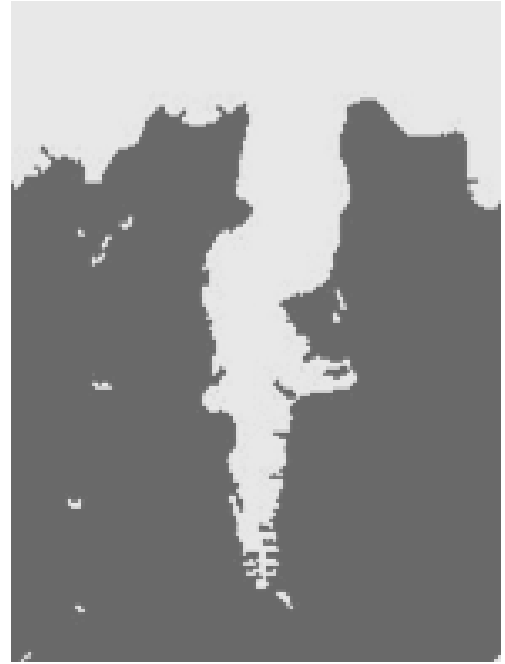

**Fig. 5.** Application of the erosion operation

**Stage 4.** Application of contour selection operators. The process of selecting contours is based on the simplest movement of the filter mask from point to point in the image; at each point (x, y), the filter response is computed using the predefined links. In the case of linear spatial filtering, the response is given by the sum of the product of the filter coefficients and the corresponding pixel values in the area covered by the filter mask. For a mask of 3x3 elements, the result (response) R of linear filtering at the point (x, y) of the image will be:

$$R = w(-1,-1)f(x-1, y-1) + w(-1,0)f(x-1, y) \cdots \\ + w(0,0)f(x, y) + \cdots w(0,0)f(x, y) + \cdots \\ w(1,0)f(x+1, y) + w(1,1)f(x+1, y+1) \quad (6)$$

Each element of the resulting matrix is the sum of the products of the mask coefficients and the pixel values directly under the mask. In particular, note that the coefficient w(0,0) stands at the value of f(x, y), thereby indicating that the mask is centered at the point (x, y).

When detecting brightness differences, discrete analogues of first and second order derivatives are used. For simplicity of presentation, one-dimensional derivatives will be considered. The first derivative of the one-dimensional function f(x) is defined as the difference between the values of neighboring elements:

$$\frac{\partial f}{\partial x} = f(x+1) - f(x) \qquad (7)$$

Here we have used partial derivative notation in order to keep the same notation in the case of two variables f(x,y), where we have to deal with partial derivatives along two spatial axes. The use of a partial derivative does not change the essence of the consideration.

Similarly, the second derivative is defined as the difference between adjacent values of the first derivative

$$\frac{\partial^2 f}{\partial x^2} = f(x+1) + f(x-1) - 2f(x) \qquad (8)$$

Since the image is being processed, which is a two-dimensional matrix, consider the two-dimensional case. The gradient is a vector function with components $\frac{\partial f}{\partial x}$ and $\frac{\partial f}{\partial y}$.

$$\nabla f = \frac{\partial f}{\partial x} dx + \frac{\partial f}{\partial y} dy \qquad (9)$$

The calculation of the first derivative of a digital image is based on various discrete approximations of the two-dimensional gradient. By definition, the gradient of the image f(x, y) at the point (x, y) is a vector of the form

$$\nabla f = \begin{bmatrix} G_x \\ G_y \end{bmatrix} = \begin{bmatrix} \frac{\partial f}{\partial x} \\ \frac{\partial f}{\partial y} \end{bmatrix} \qquad (10)$$

The direction of the gradient vector coincides with the direction of the maximum rate of change of the function f at the point (x, y). An important role in the detection of contours is played by the modulus of this vector, which is denoted $\nabla f$ and equal to:

$$\nabla f = |\nabla f| = \sqrt{G_x^2 + G_y^2} \qquad (11)$$

This value is equal to the value of the maximum rate of change of the function f at the point (x, y), and the maximum is reached in the direction of the vector $\nabla f$. The $\nabla f$ quantity is also often referred to as the gradient.

The direction of the gradient vector is also an important characteristic. Let α(x,y) denote the angle between the direction of the vector $\nabla f$ at the point (x,y) and the x-axis. Then:

$$a(x, y) = arctg\left(\frac{G_y}{G_x}\right) \qquad (12)$$

From here it is easy to find the direction of the contour at the point (x, y), which is perpendicular to the direction of the gradient vector at this point. And you can calculate the gradient of the image by calculating the values of partial derivatives $\frac{\partial f}{\partial x}$ and $\frac{\partial f}{\partial y}$ for each point [9].

**Sobel operator.** The Sobel operator uses a 3x3 image area, the principle is to use weights for the middle elements:

$$G_x = (z_7 + 2z_8 + z_9) - (z_1 + 2z_2 + z_3) \quad (13)$$

$$G_y = (z_3 + 2z_6 + z_9) - (z_1 + 2z_4 + z_5) \quad (14)$$

This increased value is used to reduce the smoothing effect by giving more weight to the midpoints [9].

The masks used by the Sobel operator have the form of matrices

$$\begin{bmatrix} 1 & -2 & 1 \\ 0 & 0 & 0 \\ 1 & 2 & 1 \end{bmatrix} \begin{bmatrix} -1 & 0 & 1 \\ -2 & 0 & 2 \\ -1 & 0 & 1 \end{bmatrix}.$$

Let the 3x3 area shown below represent the brightness values in the neighborhood of some image element

$$\begin{bmatrix} z_1 & z_2 & z_3 \\ z_4 & z_5 & z_6 \\ z_7 & z_8 & z_9 \end{bmatrix}.$$

then the result of applying the Sobel operator is shown in fig. 6.

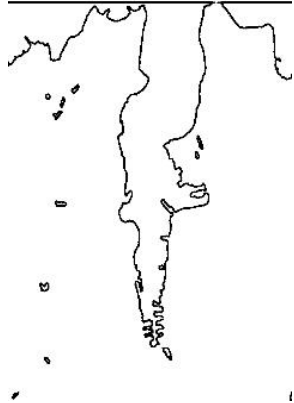

**Fig. 6.** An example of using the Sobel operator for edge selection

**Roberts operator.** One of the simplest ways to find the first partial derivatives at a point is to use the following Roberts cross gradient operator:

$$G_x = (z_9 - z_5) \quad (15)$$
$$G_y = (z_8 - z_6) \quad (16)$$

These derivatives can be implemented by processing the entire image with an operator described by masks, using a filtering procedure similar to that described above.

$$\begin{bmatrix} -1 & 0 \\ 0 & 1 \end{bmatrix} \begin{bmatrix} 0 & -1 \\ 1 & 0 \end{bmatrix}$$

Using the operator on the input image. The implementation of 2x2 masks is not very convenient, because they do not have a clearly defined central element, which significantly affects the result of filtering. But this disadvantage is compensated by a very useful property of this algorithm - the high speed of image processing [9].

The result of applying the Roberts operator is shown in Fig. 7.

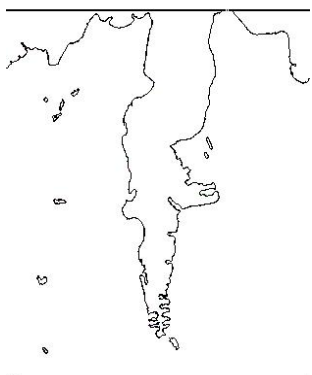

**Fig. 7.** An example of using the Roberts operator for edge selection

**Prewitt operator.** The Prewitt operator, as well as the Roberts operator, operates with a 3x3 image area, only the use of such a mask is specified by other expressions:

$$G_x = (z_7 + z_8 + z_9) - (z_1 + z_2 + z_3) \quad (17)$$

$$G_y = (z_3 + z_6 + z_9) - (z_1 + z_4 + z_7) \quad (18)$$

In these formulas, the difference between the sums along the top and bottom rows of the 3x3 neighborhood is the approximate value of the derivative along the x axis, and the difference between the sums along the first and last columns of this neighborhood is the derivative along the y axis. To implement these formulas, an operator described by masks is used, which is called the Previtt operator [9].

$$\begin{bmatrix} -1 & -1 & -1 \\ 0 & 0 & 0 \\ 1 & 1 & 1 \end{bmatrix} \begin{bmatrix} -1 & 0 & 1 \\ -1 & 0 & 1 \\ -1 & 0 & 1 \end{bmatrix}$$

The result of applying the Prewitt operator is shown in fig. eight.

In the course of the experiment, the Sobel operator proved to be the best in terms of image clarity and detail when selecting contours, the result of which is shown in Fig. 6.

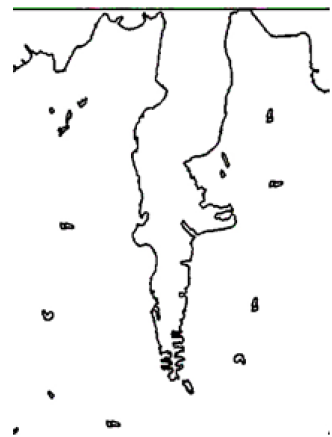

**Fig. 8.** An example of using the Prewitt operator for edge selection

Since the algorithm is used for a surface robot that will perform functions only in closed waters and coastal waters, the part that removes empty water space was used to highlight closed waters and coastal waters.

**Stage 5. Loading data from the binding file.** To calculate the GPS coordinates of the border, a method has been developed for calculating such coordinates for an arbitrary point in the image; for this, a map fragment is connected with geographical coordinates.

The SASPlanet program allows, together with a map fragment, to obtain its georeferencing file (a text file containing control points that describe the coordinate transformation). This file is used to load geographic coordinates of control points of a map fragment from the geographic coordinates file.

To increase the speed of the program, a problem-oriented algorithm for finding coordinates in the binding file is implemented. A typical method for creating a SASPlanet program binding file has been implemented and code has been written for loading lines containing data of control points, as well as converting this data into a form suitable for further work. The developed program is intended for use with fragments of maps downloaded by the SASPlanet program.

**Stage 6. Marking the GPS-coordinate area.** According to the loaded coordinates from the binding file, i.e. now we have pixel coordinates and the scale of 1 pixel (division of the pixel coordinate grid) in geo-graphic minutes is calculated for the corresponding geographical coordinates. However, since the minutes of latitude and longitude are different in geographic coordinates, the scales of 1 pixel on the latitude axis and 1 pixel on the longitude axis are calculated separately.

On fig. 9, 10 schematically show the coordinate grids used in the work and the division of the axes.

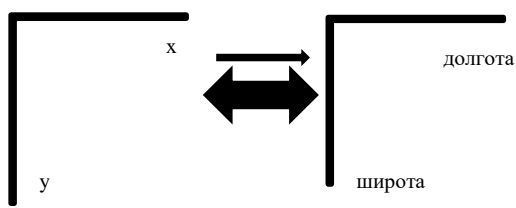

**Fig. 9**. Coordinate grids

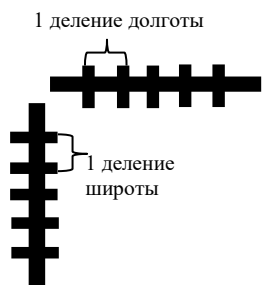

**Fig. 10.** Divisions of the coordinate axes

To calculate the division prices, the following algorithm was built and implemented as a program:

1) We translate the geographical coordinates of the corners of the coordinate grid into minutes: $C = D * 60 + M,$ where D (degrees) – degrees of geographic coordinates of the point, M (minutes) – minutes of geographic coordinates of the point, C (coordinates) – total geographic coordinates of the point in minutes.

2) Find the distance between 2 points on the coordinate line in geo-graphic minutes and in pixels

$$Cd = |K2 - K1|; Pd = |P2 - P1| \qquad (19)$$

where Cd is the distance between points in geographical minutes, Pd is the distance between points in pixels, K1 and K2 are the geographical coordinates of points 1 and 2 in minutes, P1 and P2 are the coordinates of points 1 and 2 in pixels.

3) Find the price of 1 pixel in geographic minutes: $DV = \dfrac{Cd}{Pd}$, where DV (division value) is the division price.

The final formula looks like:

$$DV = \frac{|(D2*60+M2)-(D1*60+M1)|}{|P2-P1|} \qquad (20)$$

where D1, D2 are degrees of geographical coordinates of points 1 and 2; M1, M2 – minutes of geographic coordinates of points 1 and 2; P1 and P2 – coordinates of 1 and 2 points in pixels; DV - the desired price of division.

Using this algorithm, you can find the division price along the latitude axis and the division price along the longitude axis (DVs and DVd, respectively).

**Stage 7. Obtaining GPS coordinates of the border.** After obtaining the prices of divisions of the axes of the pixel coordinate grid, it became possible to find the geographical coordinates of the boundary (contour) of the water body. Since the border is represented by an array consisting of the pixel coordinates of the points that make up the border, the calculation of the geographic coordinates of the border consists of calculating the geographic coordinates of each point in this array.

To calculate the geographical coordinates of a point, the following ratios are used:

$$Sh = Dsz * 60 + Ms - DVs * Y \qquad (21)$$

$$Dol = Ddz * 60 + Md + DVd * X \qquad (22)$$

where Sh is the desired latitude of the point, Dol is the desired longitude of the point; Dsz – degrees of latitude of the origin (of a point with pixel coordinates (0,0)), Ms – minutes of latitude of the origin; Ddz – degrees of longitude of the origin, Mh – minutes of longitude of the origin; DVs is the division value along the latitude axis, DVd is the division value along the longitude axis; Y - pixel coordinates of the point along the latitude axis, X - pixel coordinates of the point along the longitude axis.

These formulas are valid when the point of origin is in the upper left corner of the map fragment, i.e. looks like Fig. 11:

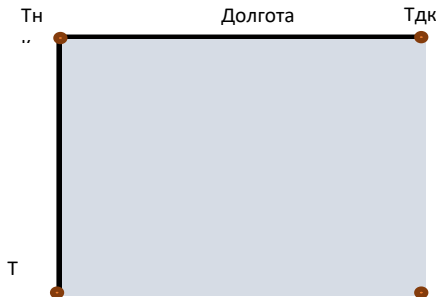

**Fig. 11.** Position of the coordinate grid

where Tnk is the origin point, Tdk is the second point along the line of longitude, Tshk is the second point along the line of latitude.

**Stage 8. Finding the area and perimeter.** Using the regionprops function, the areas and perimeters of water bodies are measured in pixels.

Then the kilometer/pixel ratio is found. To do this, the previously found DVs - the price of division along the latitude axis (ie the ratio of a geographic minute to a pixel) is multiplied by the coefficient for converting minutes into kilometers.

$KP = 1.852 * DVs$, where KP is the value of the kilometer/pixel ratio.

After that, the found areas and perimeters are converted to km using the ratio found.

$$AK = AP * KP \qquad (23)$$

$$PerK = PerP * KP \qquad (24)$$

where AP is the area in pixels, PerP is the perimeter in pixels, AK is the desired area in km, PerK is the desired perimeter in km.

**Stage 9. Saving the result.** The result of the program's work is recorded in 3 files. The area and perimeter of water bodies are recorded in the first *.xls format file. The other 2 files of *.xls and *.txt formats contain information about the GPS coordinates of the points that make up the contours of the reservoirs.

For writing, information is grouped into tables, after which it is written to files using the writable function.

**Conclusion.** This article provides an analysis of various image processing methods, on the basis of which a method for detecting boundaries for the autopilot of a surface robot is proposed. To do this, a method is used to detect certain colors in an HSV image with a predetermined water color palette. This method is the most suitable for building maps of the area from satellite images due to its data processing speed and high quality of the solution comparable to that performed manually.

To determine the water space, the methods of multispectral water indices are considered the best and most universal. However, the successful application of these methods requires a large amount of computing resources, which is unacceptable for an autopilot due to high requirements for the efficiency of decision making. Therefore, the method for determining the color in the HSV space was chosen, since all calculations for this method must be made once, during the determination of the color palette of water bodies. After that, only comparison of coordinates with the boundary file is used, without additional calculations, which increases the speed of determining water bodies.

To improve the quality of image processing, methods were used to remove noise, erosion, dilation, and edge detection.

In the course of research, the So-bel operator was chosen to implement the contour detection algorithm for the autopilot. This operator was found to be more accurate than the Roberts and Prewitt operators. In the course of research, it was found that in practice the Sobel operator detects the boundary better and more accurately, with less time than alternative options.

To make it possible to check the adequacy of the algorithm used, the program implemented a method for finding the area of water bodies. After that, the developed algorithm for detecting the GPS coordinates of the contours of water bodies is used. A positive feature of this algorithm is its high accuracy with low requirements for computing resources. The result of the algorithm execution is a file of water object boundaries in a format suitable for use by the autopilot of a surface robot.

*The study was carried out with the financial support of the Russian Foundation for Basic Research and the city of Sevastopol within the framework of the scientific project No. 18-48-920018 and within the framework of the state task of the Institute of Natural and Technical Systems No. AAAA-A19-119040590054-4.*